\documentclass[onecolumn]{aa}

\usepackage{graphicx}

\usepackage{txfonts}

\begin{document}

   \title{The importance of discovering a 3:2 twin-peak QPO in a ULX}

   \subtitle{or how to solve the puzzle of intermediate mass black holes}

   \author{
        Marek A. Abramowicz\inst{1,2,3}
        \and
        W{\l}odek Klu\'zniak\inst{1,3,4}
        \and
	Jeffrey E. McClintock\inst{5}
	\and
	Ronald A. Remillard\inst{6} }

   \offprints{W. Klu\'zniak}

   \institute{
        UKAFF supercomputer facility, Department of Physics and Astronomy, 
	University of Leicester, England
        \and
        School of Physics, Chalmers University            
	S-412-96 G\"oteborg, Sweden\\
        \email{marek@fy.chalmers.se}
        \and          
	Institute of Astronomy, Zielona G\'ora University, Wie\.za Braniborska,
        ul. Lubuska 2, PL-65-265 Zielona G\'ora, Poland
        \and
	Copernicus Astronomical Centre, ul. Bartycka 18, 00-716 Warszawa, Poland \\       \email{wlodek@camk.edu.pl}	     
	\and
        Harvard-Smithsonian Center for Astrophysics,
        60 Garden Street, Cambridge, MA 02138, U.S.A.\\
	\email{jem@cfa.harvard.edu}
	\and
	Center for Space Research, M.I.T., 
	Cambridge MA 02139, U.S.A.\\
	\email{rr@space.mit.edu}
	}

   \date{Received December, 2003; accepted }

   \abstract{ Recently, twin-peak QPOs have been observed in a 3:2
ratio for three Galactic black-hole microquasars with frequencies that
have been shown to scale as 1/$M$, as expected for general
relativisitic motion near a black hole.  It may be possible to extend
this result to distinguish between the following two disparate models
that have been proposed for the puzzling ultraluminous X-ray sources
(ULXs): (1) an intermediate-mass black hole ($ M \sim 10^3 M_{\odot}$)
radiating very near the Eddington limit and (2) a conventional black
hole ($ M \sim 10 M_{\odot}$) accreting at a highly super-Eddington
rate with its emission beamed along the rotation axis.  We suggest
that it may be possible to distinguish between these models by
detecting the counterpart of a Galactic twin-peak QPO in a ULX: the
expected frequency for the intermediate-mass black hole model is only
about 1 Hz, whereas, for the conventional black hole model the
expected frequency would be the $\sim 100$ Hz value observed for the
Galactic microquasars.  }  

\maketitle 

 \section{ULX: intermediate-mass black hole or Polish doughnut ?}

The physical nature of the recently-discovered ultraluminous X-ray sources
 (ULXs) remains a puzzle. Several of these non-nuclear, pointlike X-ray
 sources have been discovered in nearby galaxies. They have apparent
 isotropic X-ray luminosities ten to hundred times greater than the
 Eddington limit for a $10\,M_{\odot}$ black hole. The pointlike
 appearance of ULXs, together with their variability on timescales ranging
 from days to years, suggest that they must be compact accreting
 sources (Ptak \& Griffiths 1999). 

The high luminosities and compactness of ULXs were behind the claim
 that ULXs must be powered by accretion onto a new class of
 ``intermediate-mass'' black holes (Colbert \& Mushotzky 1999;
 Makishima et al. 2000). In this interpretation, ULXs are 
$\sim10^{2\pm0.5} M_{\odot}$ black holes radiating very near the
 Eddington limit. However, a different interpretation is also
 possible: a ULX could be powered by a highly super-Eddington
 accretion flow.  From works of Paczy{\'n}ski and collaborators in the
 1980s in Warsaw, it is known that thick accretion disks, which are
 supported by radiation pressure, are formed when the accretion rate
 is far above the Eddington value (Abramowicz, Jaroszy{\'n}ski \&
 Sikora, 1978; Koz{\l}owski, Abramowicz \& Jaroszy{\'n}ski,
 1978). Such a thick disk has a pair of very deep and narrow funnels
 along its accretion axis, which prompted Martin Rees  to call
 this structure a ``Polish doughnut.''  The emergent radiation flux is
 beamed in the funnels and reaches values far in excess of the
 Eddington flux (Jaroszy{\'n}ski, Abramowicz \& Paczy{\'n}ski, 1980;
 Abramowicz \& Piran 1980; Sikora 1981; Madau, 1988).  Thus, a ULX
 could very well be a Polish doughnut: i.e., an ordinary X-ray binary
 with a $\sim 10\, M_{\odot}$ black hole that is accreting at a highly
 super-Eddington rate.

\noindent Both interpretations of ULXs, the intermediate-mass black
hole and the Polish doughnut, face theoretical difficulties that
have been discussed, for example, by Strohmayer \& Mushotzky (2003)
and by King et al. (2001).  Rather than reviewing and attempting to
elaborate the uncertain arguments used in these discussions, we point
out instead an unambiguous and purely observational way to solve the
ULX puzzle.

\noindent We suggest that the mass of a ULX could be unambiguously
determined if a twin-peak QPO with frequencies in the ratio 3:2,
$\nu_{\rm upp}/\nu_{\rm lower}=1.5$, were to be discovered in the time
variability of the ULX.  Such twin-peak QPOs provide a precise mass
estimate because of the scaling found by McClintock \& Remillard
(2003) in the twin-peak kHz QPOs in microquasars,
\begin{equation}
\nu_{\rm upp} = 2.8 \,({M_{\odot}/ M}) \,{\rm kHz}.
\end{equation}

\begin{figure*}[ht]
\label{figure}
\centering
\includegraphics[angle=-90, width=120mm]{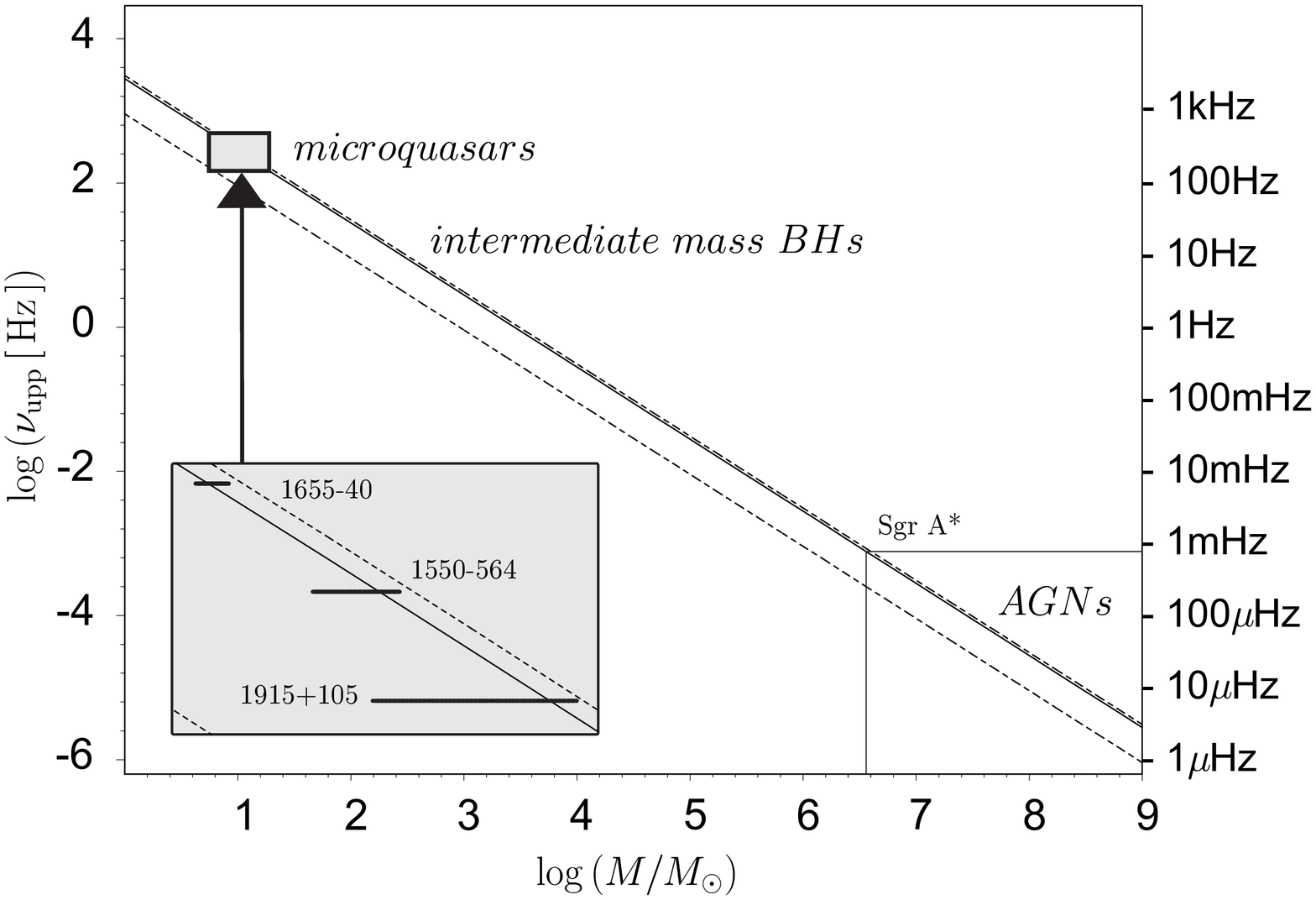}
\caption{Mirabel's microquasar-quasar analogy applied to the twin-peak
QPOs with a 3:2 frequency ratio. 
The solid line is the frequency given by eq.~[1],
the dashed (upper) and dot-dashed lines are theoretical upper and lower limits
to the upper of the twin frequencies
corresponding to dimensionless black hole spin $a=1$ and $a=-1$, respectively.
One should note that the lack of {\it a priori} knowledge of black hole spin
introduces relatively small errors in the mass estimate. The
horizontal line marked Sgr A$^*$ corresponds to a prediction 
of what the upper frequency should be for the $3.6\cdot10^6M_\odot$
 black hole at the center of our Galaxy---this frequency coincides with
that of the 17 minute flare (Genzel et al. 2003).}
\end{figure*}

\noindent The fit described by equation (1) is shown in the small insert to 
Fig.~\ref{figure}. For an extended discussion of the scaling see
also Abramowicz \& Klu\'zniak (2003); Klu\'zniak, Abramowicz,\& Lee (2003);
Abramowicz, Klu\'zniak, Stuchl{\'{\i}}k \& T\"or\"ok (2004).

\noindent The key point of our suggestion is that with this
calibration between mass and frequency in hand, Mirabel's well-known
argument (e.g., Mirabel \& Rodr\'{\i}guez, 1998) about the
similarities in black hole accretion physics over the whole
microquasar--quasar range may be used to fix the mass using the
scaling provided by (1) and shown in Fig.~\ref{figure}.


\section{QPOs}
 We stress that
not {\it all} of the high-frequency QPOs in microquasars (or neutron
star sources) scale according to (1).  Indeed, it is known that they
do not.  Thus, other ({\it non}-twin-peak) QPOs cannot be used to
reliably estimate mass. This is why the remarkable discovery of a {\it
single} QPO frequency in ULXs (Strohmayer \& Mushotzky, 2003) is not
conclusive in solving the ULX puzzle.

Many poorly understood quasi-periodicities have been observed in the
radiation flux of low-mass X-ray binaries (van der Klis 2000).  As
shown by the variability of the frequencies in a class of sources, and
even more clearly in individual sources, in general there is not a
one-to-one correspondence between the observed frequency and the
source mass.  
It is not difficult to understand why. In turbulent
accretion disks around black holes, neutron stars, and white dwarfs
most of the high frequency variability is likely connected with
transient oscillatory phenomena that occur at various locations in the
inner accretion disk. 

Because these phenomena are not uniquely
connected to any particular location that is fixed in relation
to the gravitational radius, they do not scale with $1/M$.
For example, the high-frequency QPOs vary within a few hours
or days by several hundred Hz in individual neutron-star binaries.
It is also clear that the QPOs in white dwarfs have frequencies
orders of magnitude lower than the QPOs in neutron stars,
although both classes of sources have a similar mass $M\approx 1M_\odot$
(Mauche 2002; Warner et al. 2003).
This must be related to the difference in the radii of white dwarfs
and neutron stars.
Similarly,  the correlation of low and high QPO frequencies
QPO frequencies which
spans two orders of magnitude for neutron star sources (\cite{pbk} 1999),
cannot reflect a range of stellar masses,
 but rather indicates a range of radii (and hence Keplerian frequencies)
where the QPOs are formed.

\begin{figure*}[ht]
\label{spin}
\centering
\includegraphics[angle=-90, width=90mm]{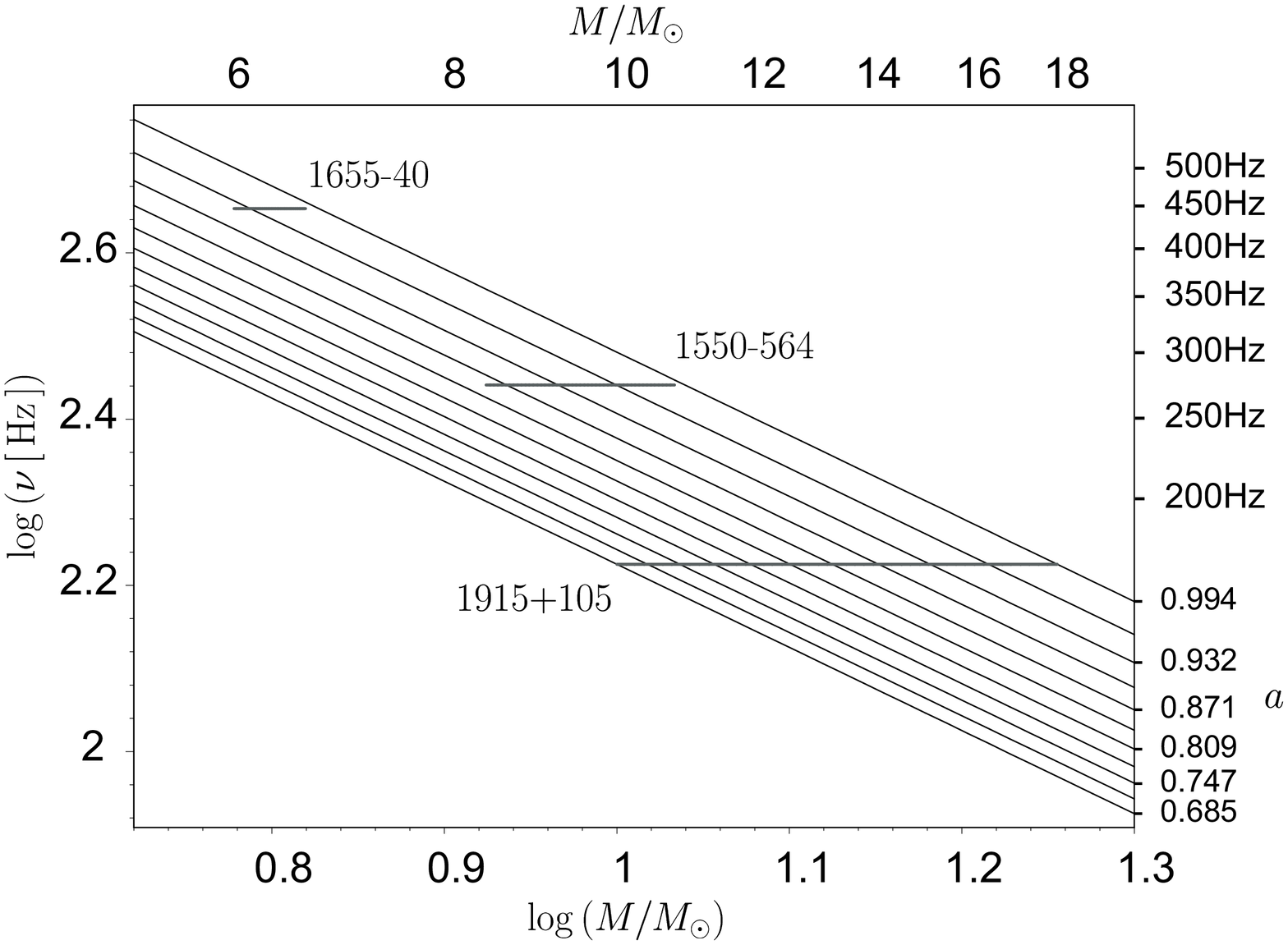}
\caption{ The 3:2 parametric resonance model for kHz twin-peak QPOs
in three microquasars. 
The observational data points from McClintock \& Remillard (2003)
are superimposed on the predictions of the parametric resonance model
of black-hole accretion disk oscillations for
various values of the spin parameter, $a$ (lower values of $a$ would give
lower frequencies).
The figure is from Abramowicz, Klu\'zniak, Stuchl{\'{\i}}k 
\& T\"or\"ok (2003).
The good agreement for reasonable values of the black hole spin shows that the
empirical fit of eq.~[1] may have a theoretical foundation, and that the
spin of the black hole may be determined if the mass is known accurately.}
\end{figure*}

\section{The twin-peak QPOs in microquasars and the 3:2 resonance}
\noindent The twin-peak QPOs in Galactic black holes are different. 
 For three such microquasars, a pair of quasi-periodic oscillations 
(twin-peak QPOs) are observed that have fairly stable frequencies
 in a 3:2 ratio and rather high
central values ($\sim 10^2\,$Hz; McClintock \&
Remillard 2003).  

Properties of
these twin-peak kHz QPOs can be understood in terms of a non-linear
resonance between the two epicyclic frequencies in the inner regions
of the accretion disc (Klu\'zniak \& Abramowicz 2000;
Klu\'zniak \& Abramowicz 2003 and references therein). In particular, the
twin-peak 3:2 QPO frequencies correspond to a specific resonance
radius that is fixed in terms of the gravitational radius of the
central compact object. This is why these frequencies scale with
$1/M$. Indeed, any general relativistic oscillation that occurs at an
orbital radius fixed in terms of $r_{\rm G} = GM/c^2$ must obviously
scale with $\nu \sim [GM/( r_{\rm G})^{3}]^{1/2} \sim 1/M$. It 
appears that observations of the three microquasars that display
twin-peak QPOs and have known masses confirm this scaling relation
(Figs.~1,2; McClintock \& Remillard 2003).

\section{Conclusions}

The detection of a twin-peak 3:2 QPO in a ULX would immediately
determine its mass (see Fig.~\ref{figure}) and thereby solve the puzzle:
is the correct model for a ULX based on an intermediate-mass black
hole ($ M \sim 10^3 M_{\odot}$) or a conventional black hole ($ M \sim
10 M_{\odot}$) embedded in a Polish doughnut?  

Finally, we note that Mirabel's quasar-microquasar analogy suggests
that one should also expect twin-peak QPOs with a 3:2 ratio in the
microhertz range for quasars 
(Fig.~\ref{figure}; $ M/M_{\odot} \sim 10^7 $ to $ 10^9$).  
In this connection, it is interesting to note that a 17
minute periodicity has recently been reported from the compact radio
source Sgr A$^*$, ``a 3.6-million-solar-mass black hole'' at the
Galactic Centre (Genzel et al. 2003). This periodicity seems to correspond
exactly to the 1 mHz frequency expected on the scaling discussed
here (Fig.~1). However, the result is inconclusive because only a single 
oscillation frequency was observed.
If the 17 minute flare period does indeed correspond to the upper (or lower)
of the twin-peak QPOs in microquasars, it would be interesting to see whether
a 26 minute (or 12 minute) quasi-periodicity may also be present in
the source.

After this paper was completed, Aschenbach et al. (astro-ph/0401589)
reported X-ray QPOs from Sgr A*. The claimed periods include 1150 s
(19 minutes) and 700s (12 minutes).

\begin{acknowledgements}
This work was supported by the UK Astrophysical Fluids Facility
(UKAFF) supported through the ``European Community --- Access to
Research Infrastructure action of the Improving Human Potential
Program'' and by the Polish KBN grant at Zielona G\'ora University. We
thank Jean-Pierre~Lasota and David Meier for comments,
and Gabriel~T\"or\"ok for drawing Figure~\ref{figure} and technical help.
\end{acknowledgements}

\end{document}